\documentclass[twoside]{article}
\pdfoutput=1

\usepackage[sc]{mathpazo} 
\usepackage[T1]{fontenc} 
\usepackage{microtype} 

\usepackage[hmarginratio=1:1,top=32mm,columnsep=20pt]{geometry} 
\usepackage{multicol} 
\usepackage[hang, small,labelfont=bf,up,textfont=it,up]{caption} 
\usepackage{booktabs} 
\usepackage{float} 

\usepackage{lettrine} 
\usepackage{paralist} 

\usepackage{abstract} 

\usepackage{titlesec} 
\renewcommand\thesection{\Roman{section}} 
\renewcommand\thesubsection{\Roman{subsection}} 
\titleformat{\section}[block]{\large\scshape\centering}{\thesection.}{1em}{} 
\titleformat{\subsection}[block]{\large}{\thesubsection.}{1em}{} 

\usepackage{fancyhdr} 
\usepackage[utf8]{inputenc}
\usepackage{listings}
\usepackage[autostyle=true]{csquotes}
\usepackage{bbm}
\usepackage[dvipsnames]{xcolor}
\usepackage[numbers]{natbib}
\usepackage{tikz}
\usepackage[hidelinks]{hyperref}

\usepackage{epsfig}
\usepackage{amssymb}
\usepackage{amsmath}
\usepackage{amsfonts}

\title{\vspace{-15mm}\fontsize{24pt}{10pt}\selectfont\textbf{Demonstrating the Feasibility of\\Automatic Game Balancing}} 

\author{
\vspace{-1cm}
\large
\textsc{Vanessa Volz}
\and
\textsc{Günter Rudolph}
\and
\textsc{Boris Naujoks}
}
\vspace{-1cm}

\date{}

\begin{document}

\maketitle 

\thispagestyle{fancy} 


\begin{abstract}
\vspace{-2mm}
\noindent   Game balancing is an important part of the (computer) game design
  process, in which designers adapt a game prototype so that the resulting
  gameplay is as entertaining as possible. In industry, the evaluation of a
  game is often based on costly playtests with human players. It suggests
  itself to automate this process using surrogate models for the prediction
  of gameplay and outcome. In this paper, the feasibility of automatic
  balancing using simulation- and deck-based objectives is investigated for
  the card game top trumps. Additionally, the necessity of a
  multi-objective approach is asserted by a comparison with the only
  known (single-objective) method. We apply a multi-objective evolutionary
  algorithm to obtain decks that optimise objectives, e.g. win rate and
  average number of tricks, developed to express the fairness and the
  excitement of a game of top trumps. The results are compared with decks
  from published top trumps decks using simulation-based objectives. The
  possibility to generate decks better or at least as good as decks from
  published top trumps decks in terms of these objectives is
  demonstrated. Our results indicate that automatic balancing with the
  presented approach is feasible even for more complex games such as
  real-time strategy games.

\end{abstract}


\begin{multicols}{2} 

\section{Introduction}
\label{sec:intro}

The increasing complexity and popularity of (computer) games result in
numerous challenges for game designers. Especially fine-tuning game
mechanics, which affects the feel and required skill profile of a game
significantly, 
is a difficult task. For example, changing the time between shots for the
sniper rifle in \emph{Halo 3} from 0.5 to 0.7 seconds impacted the gameplay
significantly according to designer Jaime
Griesemer\footnote{\url{http://www.gdcvault.com/play/1012211/Design-in-Detail-Changing-the}}.

It is important to draw attention to
the fact that the game designer's vision of a game can rarely be condensed
into just one intended game characteristic. In competitive games, for
example, it is certainly important to consider fairness, meaning
that the game outcome depends on skill rather than luck (skill-based)
and that the win rate of two equally matched players is approx. 50\%
(unbiased). But additionally, the outcome should not be deterministic and entail
exciting gameplay, possibly favouring tight outcomes.  

It therefore suggests itself to support the balancing process with tools
that can automatically evaluate and suggest different game parameter
configurations, which fulfil a set of predefined goals
(cf.~\cite{Nelson2009}). However, since the effects of certain goals tend
to be obscure at the time of design, we suggest to use a multi-objective
approach which allows to postpone the decision on a configuration until the
tradeoffs can be observed. In this paper, we introduce game balancing
as a multi-objective optimisation problem and demonstrate the feasibility
of automating the process in a case study.

For the purpose of this paper, we define game balancing as the modification
of parameters of the constitutive and operational rules of a game
(i.e. the underlying physics and the induced consequences / feedback) in
order to achieve optimal configurations in terms of a set of goals. To this
end, we analyse the card game top trumps and different approaches to
balance it automatically, demonstrating the feasibility and advantages of a
multi-objective approach as well as possibilities to introduce surrogate
models. 

In the following section, we present related work on top trumps, balancing
for multiplayer competitive games and gameplay evaluations. The subsequent
section highlights some important concepts specific to the game top trumps
and multi-objective optimisation, before the following section details our
research approach including the research questions posed. Afterwards, the
results of our analysis are presented and discussed, before we finish with
a conclusion and outlook on future work.


\section{Related Work}

\citeauthor{Cardona2014} use an evolutionary algorithm to select cards for
top trumps games from open data~\cite{Cardona2014}. The focus of their
research, however, is the potential to teach players about data and learn
about it using games. The authors develop and use a single-objective
dominance-related measure to evaluate the balance of a given deck. This
measure is used as a reference in this paper (cf. $f_D$ in
Sec. \ref{sec:approach}).

\citeauthor{Jaffe2013} introduces a technique called restricted play that
is supposed to enable designer to express balancing goals in terms of the
win rate of a suitably restricted agent~\cite{Jaffe2013}. However, this
approach necessitates expert knowledge about the game as well as an AI and
several potentially computationally expensive simulations. In contrast, we
explore other possibilities to express design goals and utilise
non-simulation based metrics.

\citeauthor{Chen2014} intend to solve \enquote{the balance problem of
  massively multiplayer online role-playing games using
  co-e\-vo\-lu\-tion\-a\-ry programming}~\cite{Chen2014}. However, they
focus on level progression and ignore any balancing concerns apart from
equalising the win-rates of different in-game characters.
 
 Yet, most work involving the evaluation of a game configuration is related
 to procedural content generation, specifically map or level generation.
 Several papers focus on issuing guarantees, e.g. with regards to
 playability \cite{Togelius2013}, solvability \cite{Smith2013}, or
 diversity \cite{Preuss2014, Isaksen2015}.
Other research areas include dynamic difficulty adaptation for
single-player games \cite{Hawkins2012}, the generation of rules
\cite{Browne2008,Smith2010b}, and more interactive versions of game
design, e.g. mixed-initiative \cite{Liapis2013a}.

\section{Basics}

In the following, the game top trumps is introduced and theoretical
background for the applied methods from multi-objective optimisation and
performance evaluation is summarised.

\subsection{Top Trumps}

\begin{figure}[H]
    \centering
    \includegraphics[width=0.6\columnwidth]{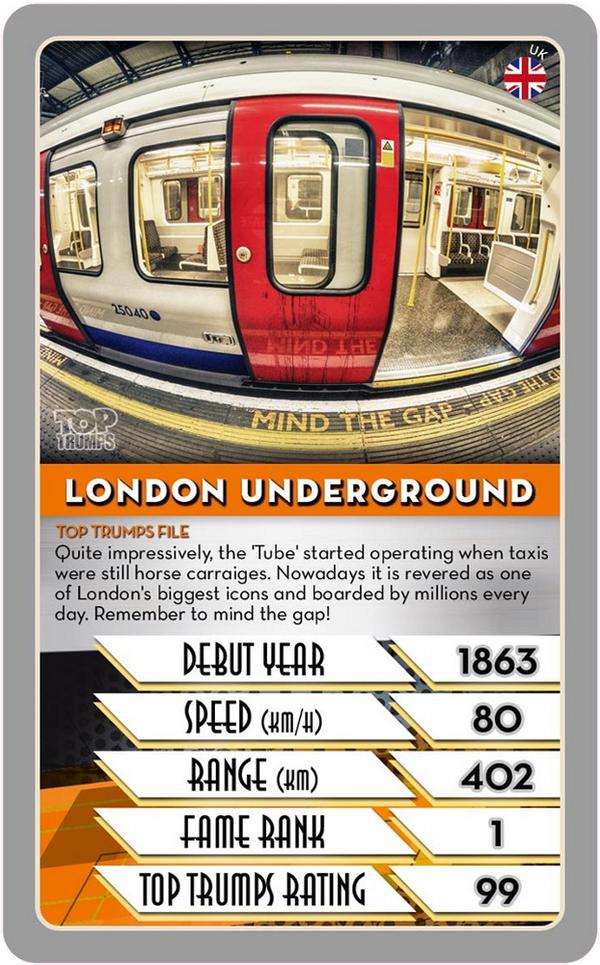}
    \caption{Card from a train-themed top trumps deck with 5 categories (\texttt{winningmoves.co.uk}) }
    \label{fig:card}
\end{figure}

Top trumps is a themed card game originally published in the 1970s and
relaunched in 1999. Popular themes include cars, motorcycles, and
aircrafts. Each card in the deck corresponds to a specific member of the
theme (such as a car model for cars) and displays several of its
characteristics, such as acceleration, cubic capacity, performance, or top
speed. An example can be found in Fig. \ref{fig:card}.

At the start of a game, the deck is shuffled and distributed evenly among
players. The starting player chooses a characteristic whose value is
then compared to the corresponding values on the cards of the remaining
players. The player with the highest value receives all cards in the trick
and then continues the game by selecting a new attribute from their next
card. The game usually ends when at least one player has lost all their
cards. 
However, for the purpose of this paper, we end the game after all
cards have been played once in order to avoid possible issues of non-ending
games.



\subsection{Multi-objective optimisation}
\label{sec:MO}

Switching from single- to multi-objective optimisation has advantages and
disadvantages~\cite{deb2001a}. While it is often better to consider
multiple objectives for technical optimisation tasks, the complete order of
individuals is lost in this case.
Instead, one has to handle incomparable solutions, e.g. two solutions that are
better than the other one in at least one objective. This
objective or component wise approach goes back to the definition of Pareto
dominance. A solution or individual $x$ is said to strictly (Pareto) dominate
another solution $y$ (denoted $x \prec y$) iff $x$ is better than $y$ in all
objectives. Considering minimisation this reads
\vspace{-2mm}
\begin{align*}
  x \prec y \quad \mbox{iff} \quad
  \forall i \in \{1, \dots , m\}: \enspace f(x_i) < f(y_i)
\end{align*}
\vspace{-2mm}
under fitness function
\begin{align*}
f : X \subset \mathbb{R}^n \to \mathbb{R}^m, f(x) = \big( f_1(x), \dots,
f_m(x) \big).
\vspace{-2mm}
\end{align*}
Based on this, the set of (Pareto) non-dominated solutions
(Pareto set) is defined as the set of incomparable solutions as defined above
and the Pareto front to be the image of the Pareto set under fitness
function $f$.

Nevertheless, even incomparable solutions need to be distinguished when it
comes to selection in an evolutionary algorithm. In the evolutionary
multi-objective optimiser considered, SMS-EMOA~\cite{bne2007a}, this is
done based on the contribution to the hypervolume (i.e the amount of
objective space covered by a Pareto front w.r.t. a predefined reference
point).
The contribution of a single solution to the overall hypervolume of the
front is used as the secondary ranking criterion for the $(\mu
+1)$-approach to selection. The first one is the non-dominated sorting rank
assigned to each solution.

For measuring the performance of different SMS-EMOA runs, we consider two
other performance indicators next to the hypervolume of the
resulting Pareto fronts. These are the additive $\varepsilon$ indicator as
well as the $R2$ indicator, all presented by Knowles et
al.~\cite{ktz2006a}. These indicators are also considered for the
termination of EMOA runs using online convergence detection as introduced
by Trautmann et al.~\cite{twsw2009a}.

For variation in SMS-EMOA, the most widely used operators in the field are
considered, namely simulated binary crossover and polynomial
mutation. These are parametrised using $p_c = 1.0, p_m = 1/n, \eta_c =
20.0,$ and $\eta_m = 15.0$, respectively, cf. Deb~\cite{deb2001a}.

\subsection{Performance Measurement for Stochastic Multi-objective Optimisers}

The unary performance indicators introduced above only express the performance of a single optimisation run. However, to evaluate and compare the relative performance of stochastic optimisers with potentially significantly different outcomes (such as evolutionary multi-objective algorithms), measurements for the \emph{statistical} performance are needed.

For this purpose, (empirical) attainment functions that describe the sets of goals
achieved with different approaches were proposed, expressing both the quality of the achieved solutions as well as their spread over multiple optimisation runs \cite{Fonseca2005}. Based on these functions, the set of goals that are reached in 50\% (or other quantiles) of the runs of the optimisers can be computed (also known as 50\%-attainment surface). Comparing the attainment surfaces of different optimisers is already a much better indicator for their performances than comparing the best solutions achieved, as they are subject to stochastic influences.

Additionally, \citeauthor{Fonseca2005} detail a statistical testing
procedure (akin to a two-sample two-sided Kolmogorov-Smirnov hypothesis
test) based on the first- and second-order at\-tain\-ment-func\-tions of two
optimisers \cite{Fonseca2005}. If the null hypotheses of these tests are
rejected, it can be assumed that the differences in performance of the
considered optimisers are statistically significant.

\section{Approach}
\label{sec:approach}
For the remainder of this paper, we denote the number of cards in a deck as
$K$ (even number) and the number of characteristics (categories) displayed
on a card $L$. Two representations are used for a deck, (1) as a vector $x \in \mathbb{R}^{K L}$ for the evolutionary algorithm and (2) as a $K \times L$ matrix $V$ for easier understanding. 

Accordingly, the value of the $k$-th card in the $l$-th
category is $v_{k,l}$ with $k \in \{1, \dots, K\}, l \in \{1, \dots,
L\}$. The $k$-th card in a deck is $v_{k,\cdot}=(v_{k,1}, \dots, v_{k,L})$. A partial order for the cards can be expressed with $v_{k_1, \cdot} \succ v_{k_2, \cdot}$ meaning that card $v_{k_2, \cdot}$ beats $v_{k_1, \cdot}$ in all categories

In this paper, we only consider decks that fulfil two basic requirements we deem existential for entertaining gameplay:
\begin{compactitem}
	\item all cards in the deck are unique:\\$\nexists (k_1,k_2) \in \{1, \dots, K\}^2$, $k_1\neq k_2$ with $ v_{k_1, \cdot} = v_{k_2, \cdot}$
	\item there is no strictly dominant card in the deck:\\$ \nexists k_1 \in \{1, \dots, K\}$ with $v_{k_2, \cdot} \prec v_{k_1, \cdot} \forall k_2 \in \{1, \dots, K\}$\\
	(in this case, dominant cards have larger values, since higher values win according to the game rules)
\end{compactitem}

We consider two agents $p_4,p_0$ with different knowledge about the played deck:
\begin{compactitem}
\item $p_4$ knows the exact values of all cards in the deck
\item $p_0$ only knows the valid value range for all values $v_{k,l}$
\end{compactitem}
Both agents are able to perfectly remember which cards have been
played already. Player $p_4$ is expected to perform better than $p_0$ on average on a balanced deck. In order to reduce the number of simulations needed to verify this, only games of a player $p_4$ against $p_0$ will be considered here.

In our simulation, both of these agents estimate the probabilities to win with each category on a given card with consideration of their respective knowledge about the deck as well as the cards already played. $p_0$ therefore has to assume a uniform distribution and will only take the values of their current card into account. $p_4$, in contrast, is able to model the probability more precisely by accounting for the number of cards with a higher value in each category that are still in play.

Let $R_G$ be the number of simulation runs. The number of tricks that $p_4$ received at the end of the $r$-th game ($r \in \{1, \dots, R_G\}$) with deck $V$ will be called $t_4^{(r,V)}$ henceforth, and thus iff $t_4^{(r,V)} > \frac{K}{2}$,
$p_4$ won the game, iff $t_4^{(r,V)} = \frac{K}{2}$ the game was a draw, and
else, $p_4$ lost. $t_c^{(r,V)}$ is the number of times the player choosing
the category did not win the trick in round $r$ of the game with deck V, i.e. the number of times the player announcing the categories changed. 

Since optimisation tasks are generally assumed to be minimisation problems (without loss of generality), this convention is satisfied here as well for the sake of consistency. Therefore, maximisation problems are transformed into minimisation task by multiplication with $-1$. In the course of this paper, we compare 8 card sets from purchased decks
with decks generated using three different approaches and corresponding fitness functions detailed in the following:
\begin{compactitem}
\item Single-objective optimisation according to the
  do\-mi\-nance-re\-la\-ted (D) measure proposed in \cite{Cardona2014}
  which describes the distance of the cards in a deck $V$ to the Pareto
  front: \vspace{-0.2cm}
  \begin{align*}
    f_D(V) = - \frac{1}{K} \sum_{k=1}^K
    \sum_{i=1}^K \left( 1- \mathbbm{1}(v_k \prec v_i) \right)
  \end{align*}
  \vspace{-0.2cm}
\item Multi-objective optimisation with si\-mu\-la\-tion-based measures developed
  with expert knowledge that are supposed to express the decks $V$'s fairness,
  excitement, and resulting balance (B):
  \vspace{-0.2cm}
  \begin{align*}
    f_B(V) = \left( -\frac{1}{R_G} \sum_{r=1}^{R_G} \mathbbm{1}\left( t_4^{(r,V)} >
        \frac{K}{2} \right), \right. \\
     \left. -\frac{1}{R_G} \sum_{r=1}^{R_G} t_c^{(r,V)}, \frac{1}{R_G}
      \sum_{r=1}^{R_G} \left|2 t_4^{(r,V)} - \frac{K}{2} \right| \right).
  \end{align*}
  \vspace{-0.2cm}
\item Multi-objective optimisation with si\-mu\-la\-tion-independent measures
  developed in the pre-experimental planning phase (cf. Sec. \ref{subsec:prePlan}) as surrogate (S) for (the
  si\-mu\-la\-tion-based) fitness $f_B$ of different decks:
  \begin{align*}
  	f_S(V) = &( -hv(V), \\ &- sd( \{
   avg(v_{\cdot,l}) | l \in \{1, \dots, L\}\})),
\end{align*}
 with the dominated hypervolume $hv$ of a deck $V$, $sd$ the empirical standard
  deviation and $avg$ the average.
\end{compactitem}

In order to compare the aforementioned approaches, an SMS-EMOA is used to
approximate the Pareto front for the fitness functions $f_S$ and
$f_B$. Online convergence detection, variation operators, and parameters as
described in Sec. \ref{sec:MO} are used. For the single-objective fitness
$f_D$, the algorithm was modified as little as possible to enable
comparisons.Thus, a $(\mu +1)$-EA was used with the same variation
operators and equivalent selection. The convergence was tested based on the
variations of some single-objective performance indicators, namely the min,
mean and max fitness values of the active population.  The experiments were
conducted using R with the help of the emoa
package\footnote{\url{https://cran.r-project.org/web/packages/emoa/}} and a
related SMS-EMOA
implementation\footnote{\url{https://github.com/olafmersmann/emoa}}${}^{,}$\footnote{Additional
  code written for the simulation and experiments will be made accessible
  after publication}.

\subsection{Research questions}
\label{sec:RQ}
The different approaches to finding a balanced top trumps deck are
evaluated and compared in Sec. \ref{sec:res}. We focus on the following
topics:
\begin{compactdesc}
\item[I] Problems of manual balancing and the solutions offered by automation
\item[II] Feasibility of automatic balancing in terms of required quality
\item[III] Performance of multi- and single-objective approaches
\item[IV] Feasibility of automatic balancing in terms of computational costs
\end{compactdesc}

\subsection{Preexperimental planning}
\label{subsec:prePlan}
Before any experiments can be executed, the test case has to be defined more
accurately. The following assumptions are made: 
\begin{compactitem}
\item The number of cards and categories are set to $K=32, L=4$ in
  accordance with these values for the purchased decks.
\item The valid range of all values $v_{l,k}$ is set to $[1,10]\in
  \mathbb{R}$, which all decks can be transformed to. This results in an
  infinite number of possible cards, but other options entail the necessity
  to construct a genotype-phenotype mapping.
\end{compactitem}

Due to the large number of possible card distributions among the players,
the order of the cards in a deck, and different starting players, a single
deck could potentially result in a large number of 
different games ($4 K!$). As a consequence, all simulation-based metrics to
evaluate the deck have to be approximated. The values of the metrics in
$f_B$, which all express an average, should be as close to the true mean of
the respective distributions as possible. To ensure the quality of the
approximation, a statistical t-test is conducted to compute the size of
the confidence interval for each metric for $R_G$ between 100 and 10\,000
at a confidence level of $0.95$. This test is repeated 500 times for each
possible sample size and each metric and. Assuming a normal distribution,
the $.95$-quantile is stored as the result. $R_G=2\,000$ games are found
to be a good tradeoff between computational time and fitness approximation
accuracy for all metrics.

An equivalent test is conducted to decide on the number of optimisation
runs necessary to approximate the performance of the corresponding approach
(to a suitable confidence interval).
Here, the $HV$-, $\epsilon$- and $R2$-indicators are considered with the
Pareto front resulting from all of these runs as a reference set.
After considering the results, the number of runs $R_O$ was set to
$R_O=100$.

For the simulation-independent approach, metrics that do not require a
simulation are developed.  The hypervolume is chosen as a measure to
achieve as many non-dominated cards in each deck as possible. This is
expected to improve the fairness of a deck (also cf. $f_D$). The standard
deviation of category means is used to increase the significance of player
$p_0$'s disadvantage, thereby resulting in a higher win-rate for $p_4$.

Different population sizes are tested and the resulting Pareto fronts are
compared for a quick estimate of their performance. Based on the results
received, approaches are evaluated on runs with population sizes of 10 and
100 individuals. This accounts for both small populations with a high
selection pressure and also bigger populations with a larger spread.

\section{Results}
\label{sec:res}

We evaluate all approaches according to the fitness function $f_B$ which is
based on expert knowledge and therefore assumed to characterise a balanced
deck best. This assumption is supported by the fact that most of the
purchased decks are located on the approximated Pareto front according to these metrics. 

In the remainder of the paper, we use the letter corresponding to the fitness function and the population size to refer to the union of the Pareto fronts of $R_O=100$ runs with the respective fitness function and population size for the multi-objective approaches. The introduced acronym with an added index $p$ refers to the Pareto front of the respective set with regards to $f_B$. A numerical index stands for the attainment surfaces to the indicated level. For example, $S10$ is the union of all Pareto fronts from optimiser runs with population size 10 and fitness function $f_S$, $S10_p$ is the Pareto front of this set and $S10_{50}$ is its 50\% attainment surface. For the single-objective approach, the union of the best individuals achieved in $R_O=100$ runs are considered instead, because the populations converge to one deck. The set of purchased decks will henceforth be denoted $PD$.

\begin{figure*}[htb]
    \centering
    \includegraphics[width=0.8\columnwidth]{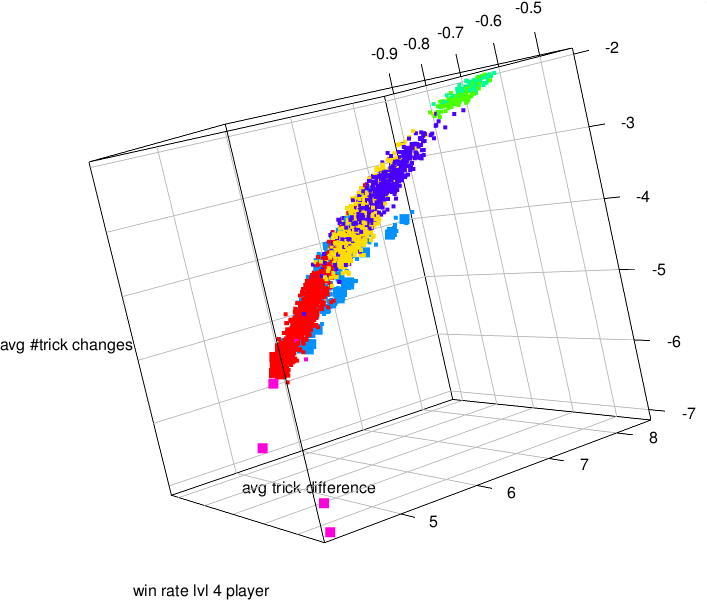}
    \hspace{2cm}
    \includegraphics[width=0.8\columnwidth]{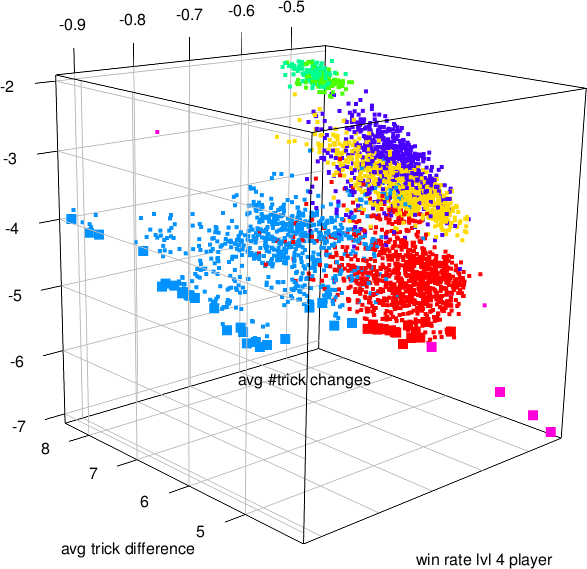}
    \caption{Union of all Pareto fronts of $R_O=100$ runs for all approaches
      considered (refer to legend in Fig.~\ref{tab:legend}) from two different perspectives. Elements of the shared Pareto front are
      depicted with larger squares.}
    \label{fig:fullAll}
\end{figure*}

\begin{figure*}[htb]
    \centering
    \includegraphics[width=0.8\columnwidth]{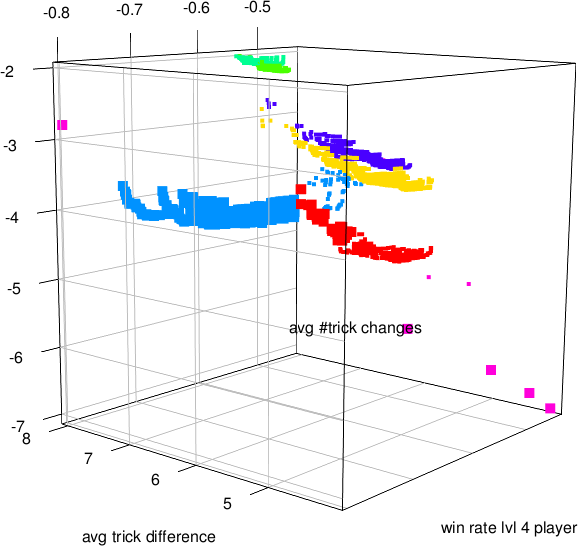}
     \hspace{2cm}
    \includegraphics[width=0.8\columnwidth]{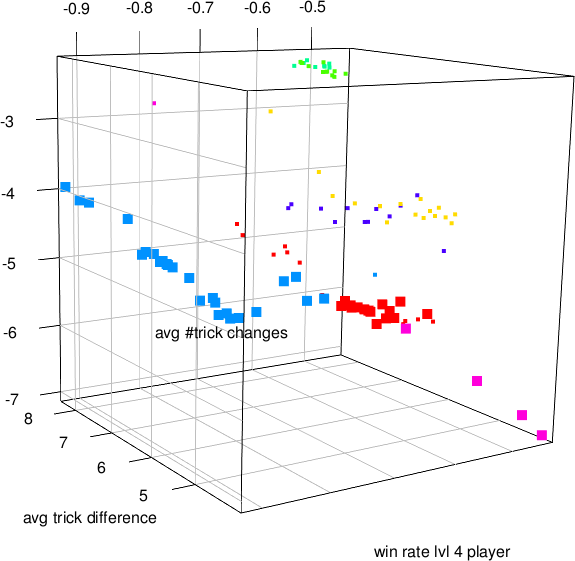}
    \caption{50\%-attainment surfaces (left) and Pareto fronts (rights) per approach of the union of all Pareto fronts from $R_O=100$ optimisation runs (depicted in Fig.~\ref{fig:fullAll}). Refer to legend in Fig.~\ref{tab:legend} for the colour scheme. Elements of the shared Pareto front are
      depicted with larger squares.}
    \label{fig:stoch}
\end{figure*}

To facilitate the discussion of the experiments, the results of the three
different approaches are plotted in terms of their performance on the
fitness function $f_B$. Figure \ref{fig:fullAll} depicts the sets listed in
Tab.~\ref{tab:legend}. Figure \ref{fig:stoch} visualises the
50\%-attainment surfaces as well as the Pareto fronts resulting from the
Pareto front union for each approach. The legend for all plots can be found
in Tab. \ref{tab:legend}, where the same colour scheme is used to refer to
the Pareto fronts and attainment surfaces of the respective approaches.

\definecolor{balancing10}{HTML}{FF0000}
\definecolor{balancing100}{HTML}{FFDB00}
\definecolor{dominance10}{HTML}{49FF00}
\definecolor{dominance100}{HTML}{00FF92}
\definecolor{surrogate10}{HTML}{0092FF}
\definecolor{surrogate100}{HTML}{4900FF}
\definecolor{realDecks}{HTML}{FF00DB}
%

\begin{table}[H]
\centering
\caption{Legend providing the color assignment to the results of different
  optimiser runs as depicted in the Fig.~\ref{fig:fullAll} and Fig.~\ref{fig:stoch}. The same colour is assigned to all results from one approach, i.e. $B10$, $B10_{50}$ and $B10_p$ have the same colour.}
\medskip
\begin{tabular}{clclclcl}
\tikz\draw[balancing10,fill=balancing10] (0,0) circle (.5ex); & B10 &
\tikz\draw[dominance10,fill=dominance10] (0,0) circle (.5ex); & D10 &
\tikz\draw[surrogate10,fill=surrogate10] (0,0) circle (.5ex); & S10 &
\tikz\draw[realDecks,fill=realDecks] (0,0) circle (.5ex); & PD \\
\tikz\draw[balancing100,fill=balancing100] (0,0) circle (.5ex); & B100 &
\tikz\draw[dominance100,fill=dominance100] (0,0) circle (.5ex); & D100 &
\tikz\draw[surrogate100,fill=surrogate100] (0,0) circle (.5ex); & S100 &\\
\end{tabular}
\label{tab:legend}
\end{table}


\subsection{Automatic Balancing Advantages}

To evaluate the advantages of automatic balancing, the following hypotheses are proposed. 
\begin{compactdesc}
\item[I-C1] The number of tests needed to approximate some si\-mu\-la\-tion-based
  metrics for a single deck to an appropriate accuracy is very high and
  possibly exceeds the number of playtests that could reasonably be done
  with human players.
\item[I-C2] Many of the purchased decks are unfair in the sense that the
  game's outcome depends strongly on luck and less on the
  players' skill levels.
\end{compactdesc}
With these hypotheses, the effort needed for manual balancing is considered and the performance of $PD$ (i.e. likely manually balanced decks) is compared to that of automatically balanced decks.

The t-test described in Sec. \ref{subsec:prePlan} already determined that
the best tradeoff between the accuracy of the approximation of
simulation-based metrics and the number of simulations $R_G$ was $\approx
2\,000$. Considering the large effort playtests with humans necessitate as
well as the bias induced by having different players play, testing 2\,000
rounds of a game with humans is tedious and potentially not even possible
on smaller budgets. For example, the standard deviation on $f_B$ for the
decks in $B10$ is $\approx (0.0427, 0.3191, 0.3576)$. If we assume we had
100 players play 10 games each, the resulting confidence interval for
$\alpha=0.05$ is $\approx (0.0442, 0.4298, 0.15)$, which would not allow
the designers to distinguish between different solutions and is therefore
not accurate enough. The standard deviation as well as the number of games
needed would likely increase with the complexity of the game as
well. Therefore, a definitive advantage of automatic balancing over manual
playtests is the possibility of a quantitative analysis of simulation-based
metrics (cf. \cite{Jaffe2013}).


Except for a single deck (motor cycles), the purchased decks are all on the
edge of the estimated Pareto front with the worst performances in terms of
the win-rate of $p_4$. This is obvious in Fig. \ref{fig:stoch}.  The low
win-rates for $p_4$ are probably due to the fact that the number of
non-dominated cards in those decks in $PD$ is relatively low. The $f_D$
average is $\approx -24.12$ compared to optimum $-31$ (only non-dominated
cards).  This means that the resulting gameplay depends heavily on luck
because there are card combinations with which a player simply cannot win
regardless of their skill. The only exception is the motor cycle deck with
a value for $f_D$ of $-30.4375$ and a much better $p_4$ win-rate of
approx. $0.8$.

Thus, we demonstrated that the effort needed to evaluate one deck
is beyond a reasonable number of play\-throughs.
Additionally, the manually balanced decks are located on the extreme edges
of the Pareto front, implying that it is difficult to find less extreme
solutions manually. This also suggests that the approximation of the Pareto
front could help a designer by giving them a more sophisticated idea about
the characteristics of their game and potential alternatives. The findings
by \citeauthor{Nelson2009} also connote that designers see potential in
automatic balancing tools to support the balancing process
\cite{Nelson2009}.

\subsection{Automatic Balancing Quality}
\label{sec:feas1}
Next, we demonstrate the feasibility of automatic balancing, i.e. that at least some of the automatically balanced decks perform on par with the purchased decks.
\begin{compactdesc}
\item[II-C1] Automatically balanced decks are on the Pareto front.
\item[II-C2] The results for II-C1 are statistically significant.
\end{compactdesc}
As is obvious from the plots (especially the right plot in
Fig.~\ref{fig:stoch}), automatically balanced decks ($S10$ and $B10$) make
up a large part of the Pareto front and are thus not dominated by the
purchased decks. Moreover, most purchased decks are concentrated at the
extreme edges of the front.

The same is true for individuals in $S10_{50}$ and $B10_{50}$, which ensures that, despite the stochasticity of the approach, decks on the Pareto front can be achieved in at least 50\% of all optimisation runs, making the result statistically relevant.

\subsection{Single- and  Multi-objective performance}
\label{sec:SOMO}
We analyse whether the multi-objectification of the approach used by
\citeauthor{Cardona2014} \cite{Cardona2014} could result in better
performing individuals. Therefore we extend $f_D$ to $f_S$. The bigger the  dominated hypervolume (the first part of $f_S$), the more non-dominated cards are in a deck, which is expressed by $f_D$. The hypervolume additionally favours cards with a larger spread, which does not affect the dominance-relationship (or the outcome of a playthrough or any si\-mu\-la\-tion-based fitness values). The second part of $f_S$ is the standard deviation of the category means. The higher the deviation, the more problematic is the strategy of player $p_0$ to assume uniform distributions for the categories to make up for their lack of knowledge.
\begin{compactdesc}
	\item[III-C1] There is a significant difference between the (empirical) attainment functions of the considered multi-objective ($S10, S100$) and single-objective ($D10, D100$) optimisation approaches.
	\item[III-C2] The results of the single-objective approach ($D10$, $D100$) perform significantly worse than the mul\-ti-ob\-jec\-tive ones ($S10$,$S100$) in terms of $f_B$.
\end{compactdesc}
In order to test hypothesis III-C1, the statistical testing procedure for
the comparison of empirical attainment surfaces described by
\citeauthor{Fonseca2005} \cite{Fonseca2005} is conducted using software
published by
C. Fonseca\footnote{\url{https://eden.dei.uc.pt/\~cmfonsec/software.html}}. With
10\,000 random mutations and $\alpha=0.05$, the null hypothesis (the
attainment function of two approaches are equally distributed) is rejected
with a p-value of 0 (critical value $0.23$, test statistic $1$) for all
comparisons in $\{D10, D100\} \times \{S10, S100\}$. This result was
expected as, judging from the visualisations (e.g. in
Fig. \ref{fig:fullAll}), the individuals found by the analysed approaches
are in very different areas. Additionally, the results found by the
single-objective approach have a very low spread, which is probably owed to
the character of the fitness measure $f_D$.


The sets of solutions found for the single-objective approaches are both
strictly dominated by both surrogate approaches according to the definition
by \citeauthor{ktz2006a} \cite{ktz2006a}. Formally, it holds that
\begin{align*}
(D10_{50}& \cup D100_{50}) \prec (D10 \cup D100) \\ 
& \prec S10_{50} \prec S10\\
(D10_{50} & \cup D100_{50}) \prec (D10 \cup D100) \\
& \| \, S100_{50} \prec
S100.
\end{align*}
The test for hypothesis III-C1 has shown that the attainment
functions of the approaches are not the same. This indicates that using
fitness function $f_S$ instead of $f_D$ has improved the results in terms
of $f_B$, thus confirming III-C2. This is suggests that the multi-objectification of $f_D$ can indeed improve the achieved results in this case.

\vspace{0.2cm}

\subsection{Computational Costs and Surrogate Objectives}
\label{sec:feasS}

We now address the feasibility of automatic balancing in terms of
computational costs. In Sec. \ref{sec:feas1}, we have already analysed and
verified its feasibility on function $f_B$. Therefore, the computational
costs needed with $R_G=2\,000$ simulations per game and $R_O=100$
optimisation runs are obviously manageable for the considered application.

However, a simulation-based approach to balancing might prove too costly
for more complex games with computationally expensive simulations or a
large game-state space. We approach this problem by investigating the
possibility of using simulation independent measures (e.g  $f_S$) instead
of $f_B$. Naturally, in practice these measures would need to be developed
in accordance with the intended balancing goals and observations of the
optimisers' behaviour, similar to what is described in
Sec. \ref{subsec:prePlan}.

The following hypotheses are put forward in order to investigate the computational costs of automatic balancing and the feasibility of simulation-independent objectives:
\begin{compactdesc}
	\item[IV-C1] Some results optimised based on fitness function $f_S$ ($S10, S100$) are not dominated by $B10$ and $B100$.
	\item[IV-C2] The best individuals in $S10$ and $S100$ perform at least equally well as the ones in $B10$ and $B100$ in terms of performance indicators.
	\item[IV-C3] There is no significant difference between the attainment functions of $S10,S100$ and $B10,B100$.
	\item[IV-C4] The results for IV-C1 and IV-C2 are statistically significant.
\end{compactdesc}
As visualised in Fig. \ref{fig:fullAll} and Fig. \ref{fig:stoch}, there are individuals in $S10$
on the shared Pareto front and which are therefore not dominated by any
individual in $B10 \cup B100$. In fact, $B100 \prec S10$ and $B10 \|
S10$. For $S100$, it can only be said that $S100 \| B100$, making IV-C1 only true for $S10$.

In order to compare the performances of the Pareto fronts of the approaches considered here, the performance indicators for HV, $\epsilon$ and R2 are computed for $S10_p$, $S100_p$, $B10_p$, $B100_p$. To facilitate the interpretation of these values, the aforementioned sets are normalised (resulting in values between 1 and 2, cf. \cite{twsw2009a}) with regards to all values achieved (cf. Fig. \ref{fig:fullAll}) before computing the indicators. The normalised Pareto front of the union of all achieved fronts is used as a reference set (cf. Fig. \ref{fig:stoch} (right)). The resulting indicator values can be found in the upper half of Tab. \ref{tab:front}. The non-dominated sorting ranks in Tab. \ref{tab:front} (top half) clearly show that hypothesis II-C2 is true for the computed values and that the approaches with the same population size perform equally well.

\begin{table}[H]
  \caption{Normalised indicator values for the Pareto fronts and 50\%-attainment surfaces of the considered approaches (cf. Fig.~\ref{fig:stoch}) as well as their resulting ranks. The rank value in brackets are alternatives accounting for
    statistically insignificant distinctions.}  
  \medskip
  \centering
\begin{tabular}{lcccl}
\hline
\textbf{set} & \textbf{HV} & $\mathbf{\epsilon}$ & \textbf{R2} & \textbf{rank}\\
\hline
$B10_p$ & 0.241 & 0.329 & 0.103 & 1\\
$B100_p$ & 0.541 & 0.559 & 0.203 & 2 (3)\\
$S10_p$ & 0.177 & 0.368 & 0.091 & 1\\
$S100_p$ & 0.475 & 0.501 & 0.206 & 2\\
\hline
$B10_{50}$ & 0.490 & 0.555 & 0.203 & 1 (1)\\
$B100_{50}$ & 0.638 & 0.650 & 0.258 & 3 (2)\\
$S10_{50}$ & 0.527 & 0.565 & 0.222 & 2 (1)\\
$S100_{50}$ & 0.676 & 0.692 & 0.288 & 4 (3)\\
\hline
\end{tabular}
 \label{tab:front}\end{table}

In order to test the statical significance of this statement, the width $w$ of the confidence interval for $\alpha=0.05$ for each set and each indicator is computed. This is done using a t-test to estimate the true indicator means on the separately achieved performance indicators for $R_O=100$ runs for each approach, normalised as before. When accounting for the uncertainty expressed in the confidence intervals, all differences in performance indicators in Tab.~\ref{tab:front} (top half) are statistically significant except for the difference in R2 for $B10$ and $S10$, as well as $B100$ and $S100$. This means that in the true ranking, $B100$ could be ranked 3 instead of 2.

The tests used to compare empirical attainment functions for hypothesis
III-C1 described in Sec. \ref{sec:SOMO} are applied again here to compare
the attainment functions for all combinations of $B10, B100$ and $S10,
S100$. Contrary to hypothesis IV-C3, the tests all reject the null hypothesis of equal attainment functions with a p-value of $0$, although the decisions are a bit tighter than in Sec. \ref{sec:SOMO}. Thus, hypothesis IV-C3 can not be confirmed. The differences in attainment functions are likely due to the fact that the compared sets occupy different areas in the objective space (cf. Fig. \ref{fig:fullAll}). This can be explained by failing to express the excitement of a playthrough in $f_S$, which was constructed to better express a deck's fairness starting from $f_D$ (cf. Sec. \ref{sec:SOMO}). Therefore, if the goal was to approximate the solutions obtained from $f_B$ with simulation-independent fitness measures, different ones should be selected, possibly using the p-value of the aforementioned test as an indicator for their quality.

From Fig. \ref{fig:stoch} (left) it is obvious that both $B10_{50}$ and
$S10_{50}$ contain individuals on the shared Pareto front, thus proving
that IV-C1 is true in at least 50\% of optimisation runs. The performance
indicators for the respective 50\%-attainment surfaces for the respective
approaches are listed in Tab. \ref{tab:front} (bottom half), along with their ranks and
the possible true rank when uncertainty is accounted for. In this case,
$S100_{50}$ performs significantly worse than all the other approaches
considered. However, $S10_{50}$ definitely performs better than $B100_{50}$
and there is no clear ranking of the performances of $S10_{50}$ and
$B10_{50}$. This implies that hypothesis IV-C4 is true as well.


Since the values in Tab.~\ref{tab:front} are all based on normalised outcomes, the absolute values can be compared. As expected, the 50\%-attainment surfaces all perform worse than their Pareto front counterpart and the differences are significant. Interestingly, the differences per indicator are smaller in the bottom half of Tab.~\ref{tab:front}. This reflects the fact that the distances of the 50\%-attainment surfaces of the different approaches in objective space are visibly smaller in Fig. \ref{fig:stoch} (left) than the Pareto fronts in Fig. \ref{fig:stoch} (right). There are also more individuals in $S100_{50}$ and $B100_{50}$ when compared to $S100_p$ and $B100_p$, respectively, which explains their smaller loss in performance indicators. This is because both approaches experience less spread in the direction of the optimum.






\subsection{Additional Observations}
\label{sec:obs}

Next to the results discussed previously, some interesting observations
were made during the experiments.

The single-objective optimisation approach converges to one deck for both
population sizes tested, even though all decks with exclusively
non-dominated cards perform equally well. The optimal fitness value for
$f_D$, 31, is achieved in almost all runs. This suggests that the algorithm
used for single-objective optimisation including the convergence detection
worked for this application. Furthermore, we can conclude that $f_D$ is not
suited for deck generation because it does not distinguish decks well. This
might be entirely different for data selection as done by \citeauthor{Cardona2014} \cite{Cardona2014}.

The optimiser runs were stopped by the convergence detection mechanism
after very different numbers of function evaluations $n_{eval}$, even for
the same approach. For example, the first 30 runs for $S10$ executed
between 3\,727 and 23\,737 fitness function evaluations. There is no
apparent correlation between $n_{eval}$ and the quality of the achieved
solutions, with $n_{eval}$ between 3\,993 and 20\,243 for runs with
solutions on the Pareto front of this subset of $S10$. This point to a high
complexity of the fitness function landscapes and validates the use of
online convergence detection in this experiment.

\section{Conclusion and Outlook}
\label{sec:con+out}

\label{sec:conclusion}
In this paper, we present our approach to automatic game balancing (as defined in Sec. \ref{sec:intro}) and apply it to the card game top trumps. Our approach includes the formalisation and interpretation of the task as a multi-objective minimisation problem which is solved using a state-of-the-art EMOA with online convergence detection. The performances of the resulting and purchased decks next to a single-objective approach \cite{Cardona2014} are evaluated using statistical analyses.

We conclusively show the feasibility of automatic game balancing in terms
of the quality of the achieved solutions for the game top trumps under the
assumptions detailed in Sec. \ref{sec:approach}. 
Being aware that computational concerns could render a simulation-based
approach infeasible for complex applications, an approach to avoid
simulation was outlined in section \ref{sec:feasS}.  The presented work,
therefore, is a necessary step to proving the feasibility of automatic
balancing in general.
Additionally, the apparent advantages of an automated balancing approach
and multi-objective balancing are discussed as well
(cf. Sec.~\ref{sec:res}). These discussions and the additional observations
in Sec.~\ref{sec:obs} strongly indicate that the presented approach was
suitable and successful.


\label{sec:outlook}


A possible way to proceed with this work is to further optimise the different parts of the approach. For example, the considered optimisers could be improved by better parameters, e.g. determined by tuning methods like sequential parameter optimisation,
thereby potentially sharpening our results. In addition, several other
modules should be tested for possible (parameter) improvements like the online
convergence detection mechanism.



With respect to the implemented player AI, it seems reasonable to extend our research by testing different improvements of the probabilistic AI used in our study. This could provide interesting results if the restriction of allowing exactly $\frac{K}{2}$ rounds of play is removed. In this case, the agent is required to plan ahead and making more complex strategies profitable. A viable AI extension is inference based reasoning about the opponent's cards as demonstrated by \citeauthor{Buro2009} in their work on improving state evaluation in trick-based card games. Monte Carlo Search is common used for card games as well, as they commonly feature imperfect information (cf. \cite{Furtak2013,Ward2009}). Another route would be the implementation of AIs that imitate human players.

Further work on the analysis of the presented measures and the discovery of
new ones is intended. As a first step in this direction, we propose to use
our approach for different applications, possibly after developing
application-specific methods. In that regard, we aim to test our approach
on more complex computer games. A first attempt will be made incorporating
The Open Racing Car Simulator
(TORCS)\footnote{\url{http://torcs.sourceforge.net/}}, but further tests on
real-time-strategy games and platformers are intended as well. Based on the
analysis of different well-performing fitness measures, a next step could
be the investigation of generalisable ones.

More importantly, we plan to evaluate our vision of a balanced deck, our fitness measures and the results of our
methods with surveys for human players. In our opinion, incorporating human
perception of balancing is the only acceptable way to achieve the eventual goal, i.e. accurately expressing and maximising human players'
enjoyment of a game.






\bibliographystyle{abbrvnat}
\bibliography{../gameBal-EMOA} 


\end{multicols}

\end{document}